\documentclass[prx,twocolumn,superscriptaddress,showpacs,amsmath,amssymb]{revtex4-1}

\usepackage{graphicx}
\usepackage{dcolumn}
\usepackage{bm}
\usepackage{times,mathptmx}
\usepackage{color}
\usepackage{multirow}
\usepackage{enumitem}
\usepackage{chngpage}

\newcommand{\ave}[1]{\langle #1 \rangle}%

\begin{document}

\title{Biphoton statistics of quantum light generated on a silicon chip} 

\author{Xiyuan Lu}
\affiliation{Department of Physics and Astronomy, University of Rochester, Rochester, NY 14627}
\author{Wei C. Jiang}
\affiliation{Institute of Optics, University of Rochester, Rochester, NY 14627}
\author{Jidong Zhang}
\affiliation{Department of Electrical and Computer Engineering, University of Rochester, Rochester, NY 14627}
\author{Qiang Lin}
\email{qiang.lin@rochester.edu}
\affiliation{Institute of Optics, University of Rochester, Rochester, NY 14627}
\affiliation{Department of Electrical and Computer Engineering, University of Rochester, Rochester, NY 14627}

\email{qiang.lin@rochester.edu}

\date{\today}


\begin{abstract}
We demonstrate a silicon-chip biphoton source with an unprecedented quantum cross correlation up to ${\rm g_{si}^{(2)}(0) = (2.58 \pm 0.16) \times 10^4}$. The emitted biphotons are intrinsically single-mode, with self correlations of ${\rm g_{ss}^{(2)}(0) = 1.90 \pm 0.05}$ and ${\rm g_{ii}^{(2)}(0) = 1.87 \pm 0.06}$ for signal and idler photons, respectively. We observe the waveform asymmetry of cross correlation between signal and idler photons and reveal the identical and non-exponential nature of self correlations of individual signal and idler photon modes, which is a nature of cavity-enhanced nonlinear optical processes. The high efficiency and high purity of the biphoton source allow us to herald single photons with a conditional self correlation $\rm g_{c}^{(2)}(0)$ as low as $\rm 0.0059 \pm 0.0014$ at a pair flux of $\rm 1.95 \times 10^5$~pairs/s, which remains below $\rm 0.026 \pm 0.001$ for a biphoton flux up to $\rm 2.93 \times 10^6$~pairs/s, with a photon preparation efficiency in the single-mode fiber up to 51\%, among the best values that have ever been reported. Our work unambiguously demonstrates that silicon photonic chips are superior material and device platforms for integrated quantum photonics.
\end{abstract}

\maketitle

\section{Introduction}
Integrated quantum photonic technology has great advances in recent years with the demonstration of diverse quantum functionalities on chip \cite{OBrien09, Walmsley15}, some of which, such as quantum simulation \cite{Walterh12} and photonic Boson sampling \cite{Walmsley13_2, White13, Walther13, Sciarrino14}, may even go beyond the reach of classical computing. Integrated photonics enable significant structural and functional complexity and thus show great promise for realizing future large-scale quantum photonic network \cite{Kimble08}. One key element underlying a fully integrated quantum photonic interconnect is to generate non-classical states of light on chip, which has attracted tremendous interest recently in demonstrating single photons \cite{Kartik12, Walmsley13, Eggleton13_2, Silberhorn13, Eggleton13, Morandotti14}, biphotons \cite{Sharping06, Takesue07, Fejer07, Baet09, Takesue10, Eggleton11, Bajoni12, Matsuda12, Eggleton12, Helmy12, Olislager13, Notomi13, Engin13, Mookherjea13, Kivshar14, Ducci14, Huang14, Peng14, Takesue14, Grassani15, Suo15, Zhu14, Thompson14, Wakabayashi15, Jiang15, Ramelow15, Savanier15}, and their quantum entanglements \cite{Takesue07, Matsuda12, Olislager13, Thompson14, Zhu14, Suo15, Wakabayashi15} on chip-scale nonlinear optical devices from various material platforms. Among the platforms developed to date, silicon is particularly attractive for quantum light generation\cite{Kartik12, Eggleton13, Morandotti14, Sharping06, Takesue07, Baet09, Takesue10, Eggleton11, Bajoni12, Matsuda12, Thompson14, Olislager13, Notomi13, Mookherjea13, Engin13, Huang14, Peng14, Takesue14, Grassani15, Suo15, Wakabayashi15, Jiang15, Savanier15}, given its mature nanofabrication technology and CMOS compatibility for high-quality device fabrication and integration, its strong optical Kerr nonlinearity for efficient nonlinear optical processes, its significant refractive index for strong mode confinement that supports device miniaturization, and particularly its clean narrowband phonon spectrum free from the broadband Raman noise \cite{Lin06} that is deleterious for quantum light sources based upon some other device platforms  \cite{Walmsley13, Eggleton13_2, Eggleton12, Kumar04, Lin072}.

In this article we demonstrate the generation of nonclassical biphoton and single-photon states in a quantum silicon photonic device with extremely high purity. We take advantage of the dramatic cavity enhancement on four-wave mixing (FWM) in a high-quality silicon microdisk resonator to produce biphotons with a signal-idler cross correlation up to $\rm g_{si}^{(2)}(0) = (2.58 \pm 0.16) \times 10^4$, the highest value reported to date \cite{Kartik12, Walmsley13, Eggleton13_2, Silberhorn13, Eggleton13, Morandotti14, Sharping06, Takesue07, Fejer07, Baet09, Takesue10, Eggleton11, Bajoni12, Matsuda12, Eggleton12, Helmy12, Olislager13, Notomi13, Engin13, Mookherjea13, Kivshar14, Ducci14, Huang14, Peng14, Takesue14, Grassani15, Suo15, Zhu14, Thompson14, Wakabayashi15, Jiang15, Ramelow15, Savanier15, Ou99, Kobayashi04, Wong06, Polzik07, Mitchell08, Guo08, Scholz09, Benson11, Harris12, Leuchs13, Riedmatten13, Thew14, Leuchs15}. Each of the signal and idler photon modes is essentially in a single mode with a photon self correlation of $\rm g_{ss}^{(2)}(0) = 1.90 \pm 0.05$ and $\rm g_{ii}^{(2)}(0) = 1.87 \pm 0.06$, respectively. The demonstrated photon cross correlation violates the classical Schwarz inequality, ${\rm g_{si}^{(2)}(0) \le [g_{ss}^{(2)}(0) g_{ii}^{(2)}(0)}]^{1/2}$, by four orders of magnitude. The biphoton production is extremely efficient, with a spectral brightness of $\rm 1.06 \times 10^9 ~ pairs/s/mW^2/GHz$, the highest value reported in current FWM-based devices \cite{Kartik12, Walmsley13, Eggleton13_2, Eggleton13, Morandotti14, Sharping06, Takesue07, Baet09, Takesue10, Eggleton11, Bajoni12, Matsuda12, Eggleton12, Engin13, Olislager13, Notomi13, Mookherjea13, Huang14, Peng14, Takesue14, Grassani15, Suo15, Wakabayashi15, Ramelow15, Jiang15, Thompson14, Kumar04}.

The strong correlation and the long lifetime of biphotons allow us to resolve the fine structure of biphoton statistics in the time domain. We not only observe the waveform asymmetry of the signal-idler cross correlation, but also experimentally reveal, for the first time to the best of our knowledge, the identical and non-exponential nature of self correlations for the signal and idler photons. All the experimental results agree very well with a theory developed for the cavity-enhanced FWM. In particular, the strong photon correlation and high efficiency of biphoton production enable heralding high-quality single photons, with conditional self correlation $\rm g_{c}^{(2)}(0)$ as low as $\rm 0.0059 \pm 0.0014$ at a pair flux of $\rm 1.95 \times 10^5$~pairs/s. $\rm g_{c}^{(2)}(0)$ remains as low as $\rm 0.026 \pm 0.001$ even at a biphoton flux as large as $\rm 2.93 \times 10^6$~pairs/s, with a preparation efficiency up to 51\% for the photons collected directly in the single-mode fiber. The recorded $\rm g_{c}^{(2)}(0)$ values are among the best that have ever been reported \cite{Walmsley13, Silberhorn13}.
\begin{figure}[t!]
\centering\includegraphics[width=1.0\columnwidth]{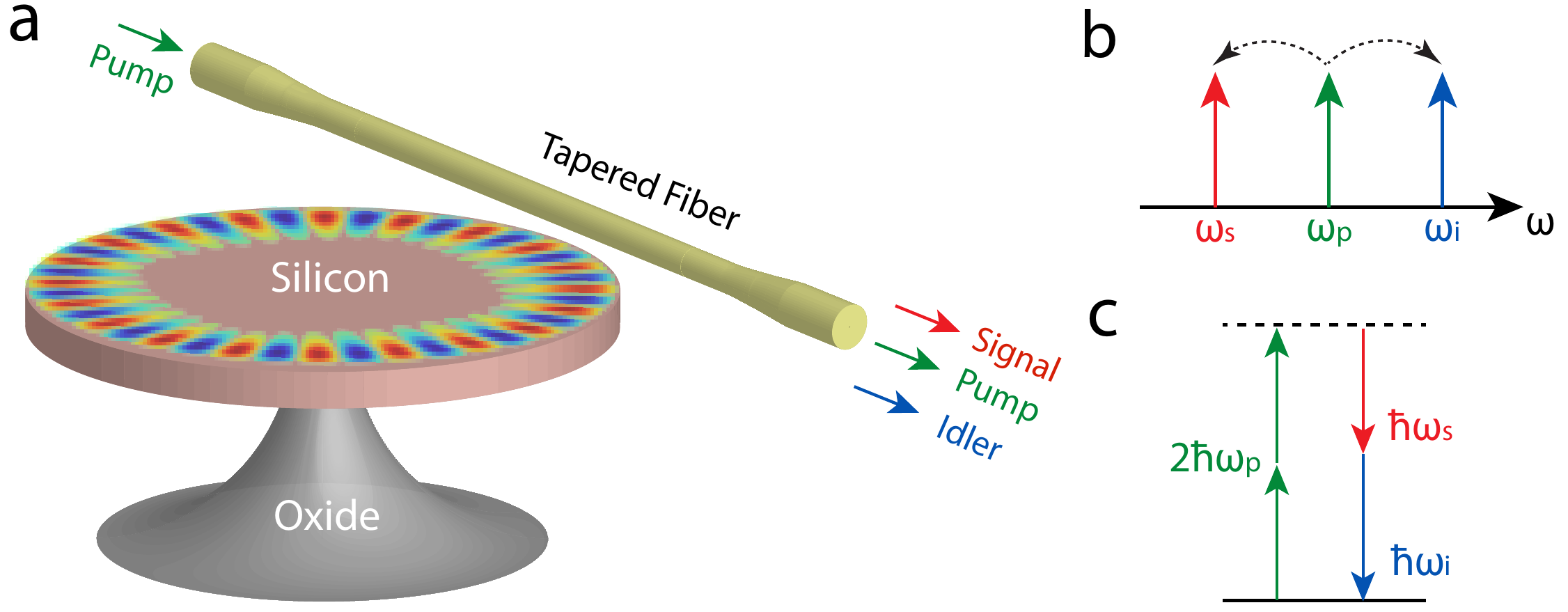}
\caption{\label{Scheme} {\bf Scheme of the photon source.} (a) A schematic illustration of the photon-pair generation process. The photons are coupled into and out of the device by the same tapered single-mode fiber. (b) and (c) show the frequency diagram and energy diagram, respectively, for the cavity-enhanced four-wave mixing process.}
\end{figure}

\section{Device and Setup}
The device is a high-Q silicon microdisk resonator with a diameter of 9~$\mu$m and a thickness of 260~nm (Fig.~\ref{Scheme}(a)), sitting on the silica pedestal. The employed device structure enables precise dispersion engineering \cite{Jiang15} to satisfy the frequency matching condition among three cavity modes involved in the FWM process. As a result, pumping at the central mode produces biphotons in signal and idler modes with frequencies located symmetrically around the pump, as shown in Fig.~\ref{Scheme}(b,c). The pump wave is coupled into the device by a tapered single-mode optical fiber, which also delivers the produced biphotons out of the device. An excellent feature of such coupling scheme is that photons are delivered to the same single spatial mode of the coupling fiber for all the optical modes involved in the FWM process. Moreover, the fiber-device coupling scheme enables flexible engineering of the photon coupling efficiency simply by changing the device-taper distance.
\begin{figure*}[htbp]
\centering\includegraphics[width=1.8\columnwidth]{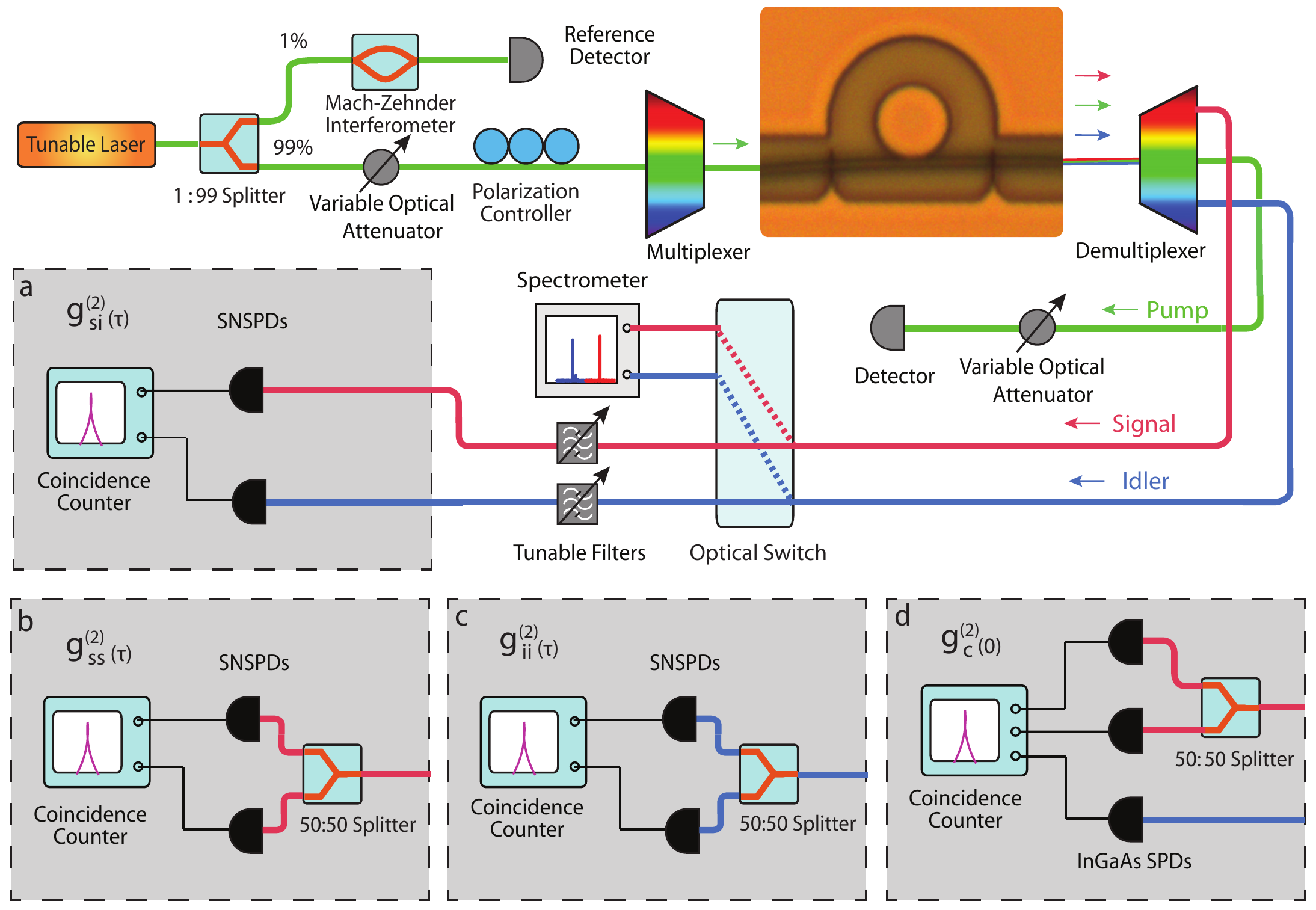}
\caption{\label{Setup} {\bf Experimental setup.} Setup (a) measures the cross correlation between signal and idler. Setups (b) and (c) measure the self correlations of individual signal and idler modes, respectively. Setup (d) characterizes the antibunching of heralded single photons by conditional self correlation. SNSPD: niobium nitride (NbN) superconducting nanowire single-photon detector, with a quantum efficiency of 7\% and a timing jitter of 70~ps. InGaAs SPD: indium gallium arsenide single-photon detector, with a quantum efficiency of 15\% and a timing jitter around 250~ps,. }
\end{figure*}

The fabricated device is tested with the experimental setup shown in Fig.~\ref{Setup}. To filter out the laser noise, the pump laser passes through a coarse wavelength-division multiplexer before coupled into the device. The multiplexer has a 3-dB bandwidth of 17~nm for each of its transmission bands whose center wavelengths are separated by 20~nm apart with a band isolation over 120~dB. The biphotons generated from the device are then separated into individual photon modes by a demultiplexer, which is identical to the multiplexer used at the input end. The photoluminescence spectra of the biphotons are recorded at each transmission port of the demultiplexer to suppress the pump wave. Tunable bandpass filters with a 3-dB bandwidth of 1.2~nm are used in front of the single-photon detectors to cut the Raman noises produced in the delivery silica fibers \cite{Jiang15}. The quantum correlation properties of the photons are characterized by single photon detectors and coincidence counting setups shown in the insets (a-d) of Fig.~\ref{Setup}.

\section{Biphoton generation}
\begin{figure}[t!]
\centering\includegraphics[width=1.0\columnwidth]{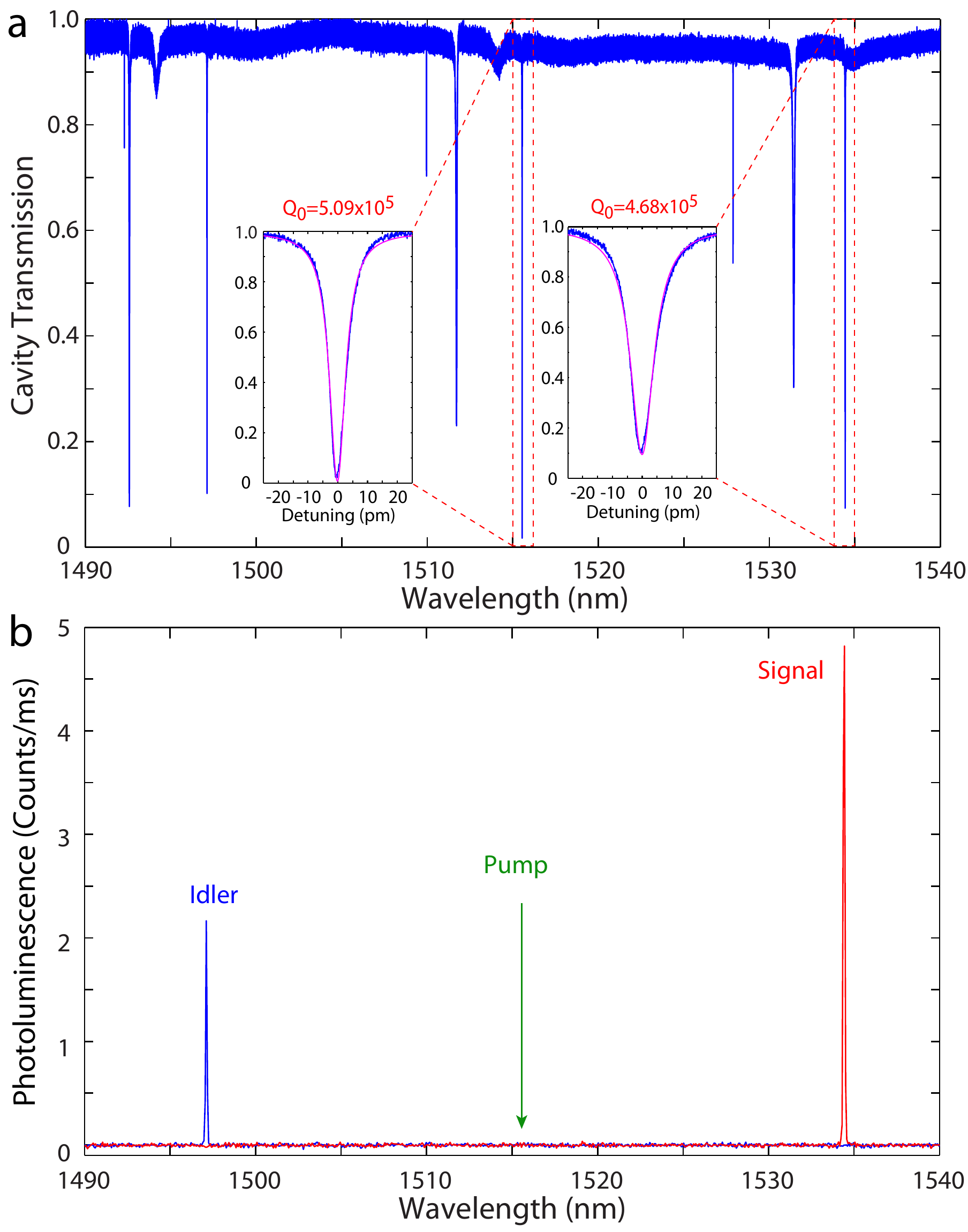}
\caption{\label{PL} {\bf Cavity transmission and photoluminescence spectra.} (a) The transmission spectrum of the passive cavity scanned by a tunable continuous-wave laser. The insets show detailed transmission spectra, with theoretical fitting shown in red. (b) The photoluminescence spectra of the generated biphotons. The colors indicate the individual spectra recorded at different transmission ports of the demultiplexer. The spectrometer has a resolution of $\sim$16 GHz.}
\end{figure}

The laser-scanned transmission spectrum of the passive cavity (Fig.~\ref{PL}(a)) shows that a constant mode spacing of 2.437~THz is achieved for the cavity modes located at 1497.1, 1515.6, and 1534.4~nm, with intrinsic Q factors of ${\rm 3.33 \times 10^5}$, ${\rm 5.09 \times 10^5}$, and ${\rm 4.68 \times 10^5}$, respectively. Consequently, a pair of clean photon modes is produced by pumping at the central mode at 1515.6~nm (Fig.~\ref{PL}(b)). The spectra show no noise mode at all, demonstrating an excellent spectral cleanness of the cavity-enhanced FWM process in silicon. The amplitude difference between the two photon modes is due to different external couplings of cavity modes to the tapered fiber. 

To characterize the biphoton source, we record the biphoton flux when pumping the device with a continuous-wave (CW) laser, as shown in Fig.~\ref{Flux}. The device is operating under a condition close to critical coupling for optimized power efficiency, with loaded Q factors of ${\rm 2.23 \times 10^5}$, ${\rm 2.97 \times 10^5}$, and ${\rm 2.59 \times 10^5}$ for idler, pump, and signal modes, respectively. The biphoton flux depends quadratically on the pump power, which is a characteristic of the degenerate FWM process. A small pump power of ${\rm 53.8~\mu W}$ dropped into the cavity generates a biphoton flux of ${\rm 1.61\times 10^6~pairs/s}$, clearly demonstrating the high efficiency of the device. Such a high efficiency results from the significant Purcell enhancement on the triply resonant degenerate FWM process inside the cavity \cite{Jiang15}, with a biphoton emission rate of ${\rm R_{si} =2 (g N_p)^2 \Gamma_{es} \Gamma_{ei}/(\bar{\Gamma} \Gamma_{ts} \Gamma_{ti})}$ where $N_p$ represents the pump photons inside the cavity, ${\rm \Gamma_{ej}}$ and ${\rm \Gamma_{tj}}$ are the external coupling rate and the photon decay rate of the loaded cavity, with subscripts ${\rm j = s,i}$ denote the signal and idler modes, respectively, and ${\rm \bar{\Gamma}=(\Gamma_{\rm ts} + \Gamma_{\rm ti})/2}$. $g$ represents the vacuum coupling rate of the FWM process (see Appendix). FWM creates biphotons inside the cavity at a rate of ${\rm 2 (g N_p)^2/{\bar{\Gamma}}}$, which are delivered to the coupling waveguide with a pair extraction efficiency of ${\rm \Gamma_{es} \Gamma_{ei}/(\Gamma_{ts} \Gamma_{ti})}$. As shown in Fig.~\ref{Flux}, the theoretical prediction of biphoton flux agrees closely with the experimental observation.

The generation efficiency of biphotons is characterized by spectral brightness, the biphoton flux per unit spectral width per unit pump power square, since degenerate FWM depends quadratically on the pump power. Detailed analysis shows that the emitted signal and idler photons exhibit identical emission spectra profile centering around cavity resonance frequencies (\emph{Appendix}),
\begin{eqnarray}
S_{\rm j}(\omega) \propto \frac{g^2 N_p^2}{[(\omega-\omega_{\rm j})^2+(\Gamma_{\rm ts}/2)^2][(\omega-\omega_{\rm j})^2+(\Gamma_{\rm ti}/2)^2]}, \label{Eq_Spectra}
\end{eqnarray}
where j=s,i stands for signal or idler mode and $\rm \omega_j$ represents the center frequency of the cavity resonance. Eq.~(\ref{Eq_Spectra}) shows that the biphoton emission spectra exhibit a spectral width narrower than that of the passive cavity modes. The loaded cavity linewidths of signal and idler modes are 0.755~GHz and 0.899~GHz, respectively, which lead to a spectral linewidth of 0.527~GHz for the emitted biphotons. The spectral brightness of the emitted biphotons is thus inferred to be ${\rm 1.06\times 10^9~pairs/s/mW^2/GHz}$. This value is the highest in FWM-based photon-pair sources reported up to date \cite{Kartik12, Walmsley13, Eggleton13_2, Eggleton13, Morandotti14, Sharping06, Takesue07, Baet09, Takesue10, Eggleton11, Bajoni12, Matsuda12, Eggleton12, Engin13, Olislager13, Notomi13, Mookherjea13, Huang14, Peng14, Takesue14, Grassani15, Suo15, Wakabayashi15, Ramelow15, Jiang15, Thompson14, Kumar04}. It is one order of magnitude higher than our previous device \cite{Jiang15}, since the current device exhibits clean singlet cavity resonances (Fig.~\ref{PL}(a)) that are free from the photon backscattering. 

\section{Cross Correlation}
The quantum cross correlation between the signal and idler photons is characterized by ${\rm g_{si}^{(2)}(\tau) \equiv \frac{\ave{E_i^\dag(t) E_s^\dag(t+\tau) E_s(t+\tau) E_i(t)}}{\ave{E_i^\dag(t)E_i(t)} \ave{E_s^\dag(t)E_s(t)}}}$, where ${\rm E_j(t)}$ (${\rm j = s,i}$) is the field operator for the signal and idler modes \cite{Mandelbook}. For the cavity enhanced FWM process, detailed analysis shows that the emitted biphotons exhibit the following cross correlation (\emph{Appendix})
\begin{eqnarray}
g_{\rm si}^{(2)}(\tau) =
  \begin{cases}
    \frac{\Gamma_{\rm ts} \Gamma_{\rm ti}}{4 (g N_{\rm p})^2} e^{-\Gamma_{\rm ts} \tau} + 1 & \quad (\tau \ge 0), \\
    \frac{\Gamma_{\rm ts} \Gamma_{\rm ti}}{4 (g N_{\rm p})^2} e^{\Gamma_{\rm ti} \tau}  + 1 & \quad (\tau < 0).
  \end{cases} \label{Eq_CrossCor}
\end{eqnarray}
with a peak value of ${\rm g_{si}^{(2)}(0) = \Gamma_{ts} \Gamma_{ti}/ \left[ 4 (g N_p)^2\right] + 1}$, which scales inversely with the power square of the pumping level. When the pump power is weak (${\rm gN_p \ll \Gamma_{ts,ti}}$), the generated biphotons are in the single-photon regime. When the pump power is strong (${\rm gN_p \gg \Gamma_{ts,ti}}$), the source falls into the regime dominated by multi-photon generation.

\begin{figure}[t!]
\centering\includegraphics[width=1.0\columnwidth]{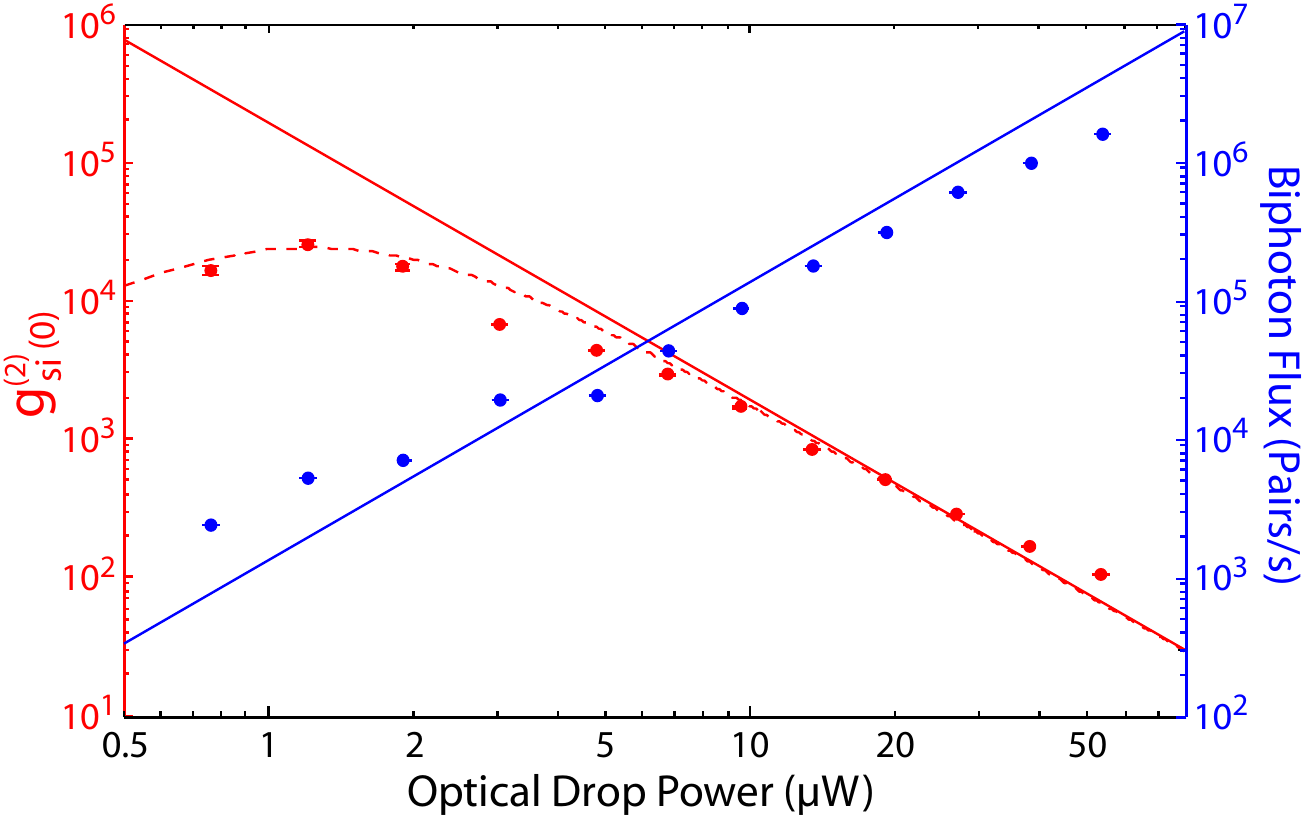}
\caption{\label{Flux} {\bf Photon pair flux (blue) and the cross-correlation $g_{\rm si}^{(2)}(0)$ (red) as a function of pump power dropped into the cavity.} The experimental data are shown in dots and the solid curves are theoretical predictions. The dashed red line takes into account the detector dark counts and the Raman noises produced in the delivery fibers.
}
\end{figure}

Figure~\ref{Flux} shows the power dependence of cross correlation (${\rm g_{si}^{(2)}(0)}$), biphoton flux (${\rm R_{si}}$) and their theoretical predictions. The produced biphotons exhibit a high cross correlation, with ${\rm g_{si}^{(2)}(0) = 103 \pm 0.2}$ at a biphoton flux of ${\rm 1.61 \times 10^6~pairs/s}$. ${\rm g_{si}^{(2)}(0)}$ increases with decreased pair flux, following closely the theoretical expectation in the red solid curve. The cross correlation reaches a peak value of ${\rm (2.58 \pm 0.16) \times 10^4}$ at a biphoton flux of ${\rm 5.25 \times 10^3~pairs/s}$. This quantum cross correlation suggests that among over 25,000 detected photon pairs, only one pair is not correlated with each other. This value is the highest one reported to date among all biphoton sources \cite{Kartik12, Walmsley13, Eggleton13_2, Silberhorn13, Eggleton13, Morandotti14, Sharping06, Takesue07, Fejer07, Baet09, Takesue10, Eggleton11, Bajoni12, Matsuda12, Eggleton12, Helmy12, Olislager13, Notomi13, Engin13, Mookherjea13, Kivshar14, Ducci14, Huang14, Peng14, Takesue14, Grassani15, Suo15, Zhu14, Thompson14, Wakabayashi15, Jiang15, Ramelow15, Savanier15, Ou99, Kobayashi04, Wong06, Polzik07, Mitchell08, Guo08, Scholz09, Benson11, Harris12, Leuchs13, Riedmatten13, Thew14, Leuchs15}. As biphoton flux decreases, the recorded ${\rm g_{si}^{(2)}(0)}$ deviates from the theoretical prediction, which is mainly due to the detector dark counts and the Raman noise from the optical fibers. Taking into account these two factors, the prediction in dashed red line agrees with experimental data quite well (Fig.~\ref{Flux}).

The long cavity lifetime of the device allows us to temporally resolve the quantum cross correlation. Two examples are shown in Fig.~\ref{CrossCorrelation}(a,b) with pump powers of 1.2 and 53.8 ${\rm \mu W}$, respectively. The temporal waveforms of cross correlations are consistent with the theoretical predictions. Note that ${\rm g_{si}^{(2)}(\tau)}$ becomes asymmetric with respect to ${\rm \tau}$ when the signal and idler photons exhibit different lifetimes in the cavity. Although biphotons are created simultaneously, individual signal (or idler) photons stay inside the cavity over their photon lifetimes before transmitting out of the cavity. Consequently, the signal (or idler) photon transmitted later are correlated with the idler (or signal) photon transmitted earlier, within the lifetime of the signal (or idler) photon. The asymmetric exponential decay is a fingerprint of the cavity-enhanced nonlinear optical process \cite{Scholz09}, which is distinct from the waveguide-based biphoton sources \cite{Sharping06, Takesue07, Baet09, Takesue10, Eggleton11, Matsuda12, Eggleton12, Helmy12, Thompson14, Olislager13, Mookherjea13, Huang14}.

This feature is clearly observed in Fig.~\ref{CrossCorrelation}(c). We measured ${\rm g_{si}^{(2)}(\tau)}$ and ${\rm g_{is}^{(2)}(\tau)}$ by switching the signal and idler paths before the SSPDs in Fig.~\ref{Setup}(a). The tapered fiber coupling scheme leads to different photon lifetimes of the two modes and thus enables the observation of the asymmetric waveform of $g_{\rm si}^{(2)}(\tau)$. Fig.~\ref{CrossCorrelation}(c) shows a clear asymmetric $g_{\rm si}^{(2)}(\tau)$ resulting from different lifetimes of 98.5 and 158 ps for the signal and idler photons, respectively. To the best of our knowledge, this is the first time to observe such an asymmetric cross correlation for a chip-scale photon-pair source \cite{Kartik12, Walmsley13, Eggleton13_2, Silberhorn13, Eggleton13, Engin13, Morandotti14, Sharping06, Takesue07, Baet09, Takesue10, Eggleton11, Bajoni12, Matsuda12, Eggleton12, Helmy12, Thompson14, Olislager13, Notomi13, Mookherjea13, Kivshar14, Ducci14, Huang14, Peng14, Takesue14, Grassani15, Suo15, Wakabayashi15, Jiang15, Zhu14}. 

\section{Self Correlation}
\begin{figure}[t!]
\centering\includegraphics[width=1.0\columnwidth]{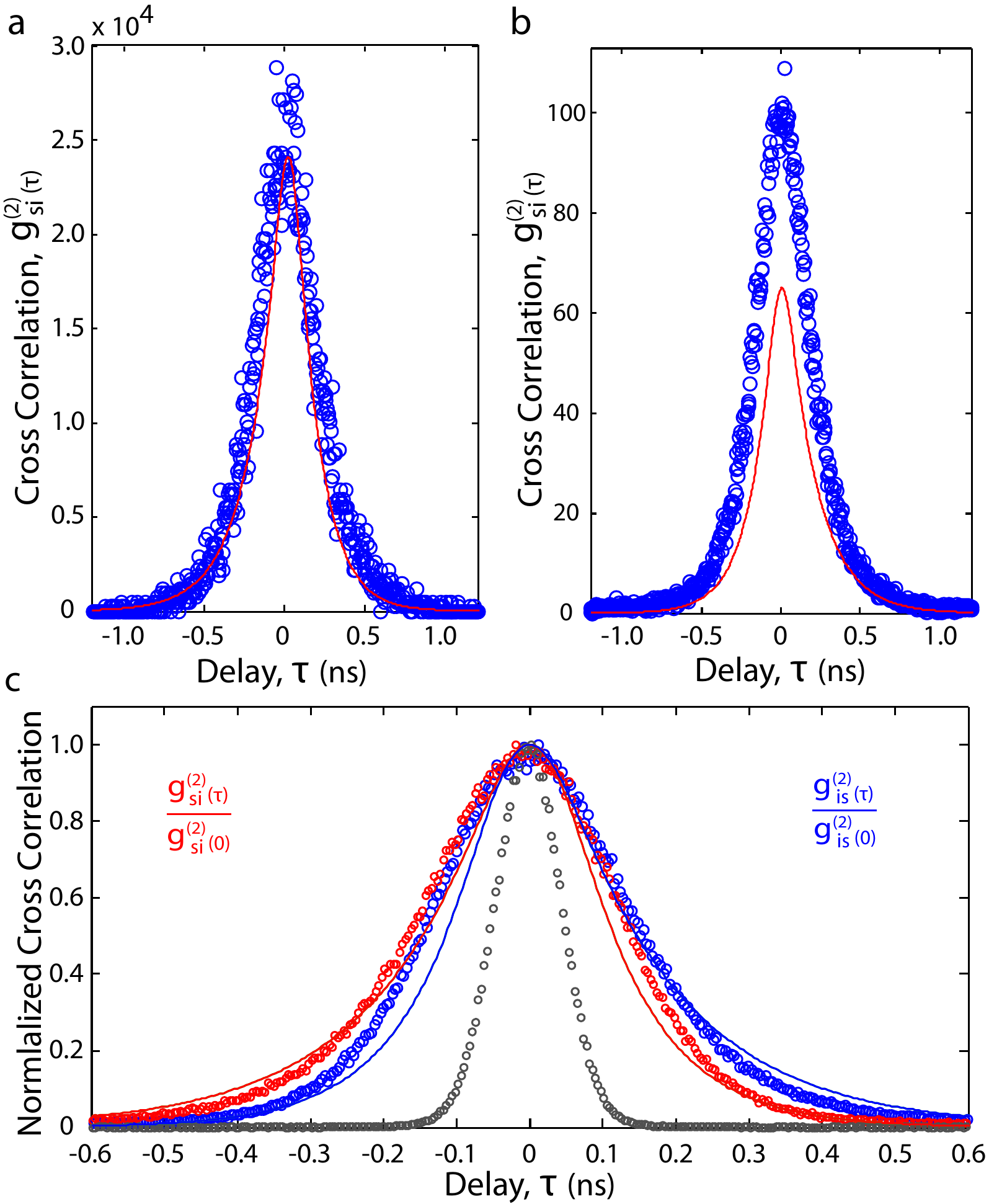}
\caption{\label{CrossCorrelation} {\bf Cross correlations between signal and idler photons.} (a) and (b) show $g_{\rm si}^{(2)}(\tau)$ at two optical drop powers of 1.2 and 53.8~$\mu$W, respectively, with experimental data in blue and the theoretical prediction in red. (c) Normalized cross correlations, $g_{\rm si}^{(2)}(\tau)/g_{\rm si}^{(2)}(0)$ and $g_{\rm is}^{(2)}(\tau)/g_{\rm is}^{(2)}(0)$, with experimental data shown in dots and theory in solid curves. The loaded cavity qualities are ${\rm 1.21 \times 10^5}$ and ${\rm 1.99 \times 10^5}$ for signal and idler modes, respectively. The gray dots show the recorded instrument response of the detection system, which is dominated by the timing jitters of the two SSPDs. The cross correlations are recorded with a time bin of 4~ps.}
\end{figure}

\begin{figure}[t!]
\centering\includegraphics[width=1.0\columnwidth]{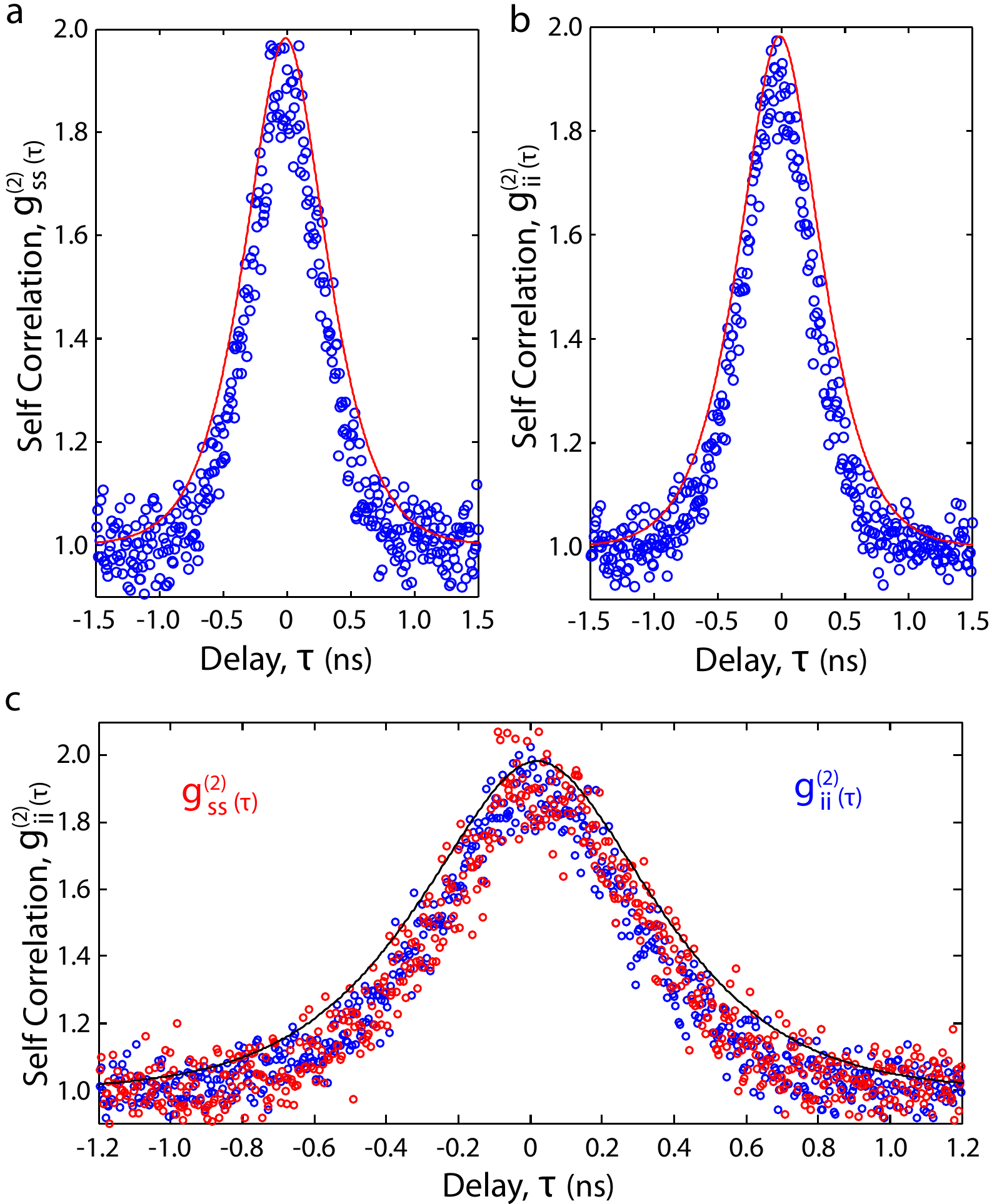}
\caption{\label{SelfCorrelation} {\bf Self correlations of signal and idler photons.} (a) and (b) show the self-correlation functions of signal and idler photons, respectively, with experimental data shown in blue and theory in red. (c) Direct comparison between $g_{\rm ss}^{(2)}(\tau)$ (blue) and $g_{\rm ii}^{(2)}(\tau)$ (red), together with the theory in black.  The theoretical curves in (a)-(c) are obtained from Eq.~(\ref{Eq_SelfCor}) convolved by the instrument response (Fig.~\ref{CrossCorrelation}(c), gray curve). The self-correlation functions are recorded with a time bin size of 8~ps.}
\end{figure}

The self correlations of the signal and idler photons are characterized by the following expression, $g_{\rm jj}^{(2)}(\tau) \equiv \frac{\ave{E_j^\dag(t) E_j^\dag(t+\tau) E_j(t+\tau) E_j(t)}} {\ave{E_j^\dag(t)E_j(t)}^2 }$ ($j=s,i$) \cite{Mandelbook}. In general, $g_{\rm jj}^{(2)}(0)$ is a measure of the mode nature of the produced photons, $g_{\rm jj}^{(2)}(0)=1 + 1/K$ where $K$ is the number of Schmidt modes the photons occupy \cite{Eberly06}. For an ideal single-mode ($K=1$) thermal source, $g_{\rm jj}^{(2)}(0)=2$. Our analysis shows that the signal and idler photons produced from the cavity enhanced FWM process exhibit the following self correlation (\emph{Appendix})
\begin{eqnarray}
 g_{\rm jj}^{(2)}(\tau) = \left[ \frac{\Gamma_{\rm ts} e^{-\Gamma_{\rm ti}|\tau|/2} - \Gamma_{\rm ti} e^{-\Gamma_{\rm ts}|\tau|/2} }{\Gamma_{\rm ts} - \Gamma_{\rm ti}}\right]^2 +1, \label{Eq_SelfCor}
\end{eqnarray}
where $j=s,i$ stands for the signal and idler photons, respectively. Equation (\ref{Eq_SelfCor}) shows $g_{\rm jj}^{(2)}(0)=2$. In practice, the finite instrument response of the detection system (Fig.~\ref{CrossCorrelation}c, gray curve) will slightly broaden the self correlation function, leading to a slightly reduced peak value of 1.98, as shown by the solid curve in Fig.~\ref{SelfCorrelation}.

Figure~\ref{SelfCorrelation}(a) and (b) show the recorded self-correlation functions (blue dots) for signal and idler photons, respectively, which agree well with the theoretical prediction (red curve). These measurements show $g_{\rm ss}^{(2)}(0) = 1.90 \pm 0.05$ and $g_{\rm ii}^{(2)}(0) = 1.87 \pm 0.06$ for the signal and idler, respectively, which are very close to the theoretical value of 1.98, and thus reveal the single mode nature of the produced photons. This single mode nature of the photons results from the high finesse of the device, which is distinctive from the cavity-enhanced SPDC that generally produces photons in multi-modes \cite{Ou99, Kobayashi04, Wong06, Polzik07, Mitchell08, Guo08, Scholz09, Benson11, Harris12, Leuchs13, Riedmatten13, Thew14, Leuchs15}. Moreover, the recorded values of $g_{\rm ss}^{(2)}(0)$, $g_{\rm ii}^{(2)}(0)$, and $g_{\rm si}^{(2)}(0)$ violate the classical Schwarz inequality \cite{Mandelbook}, ${\rm g_{si}^{(2)}(0) \le [g_{ss}^{(2)}(0) g_{ii}^{(2)}(0)}]^{1/2}$, by more than four orders of magnitude, which clearly shows the nonclassical nature of the biphotons generated by the device.

\begin{figure}[t!]
\centering\includegraphics[width=1.0\columnwidth]{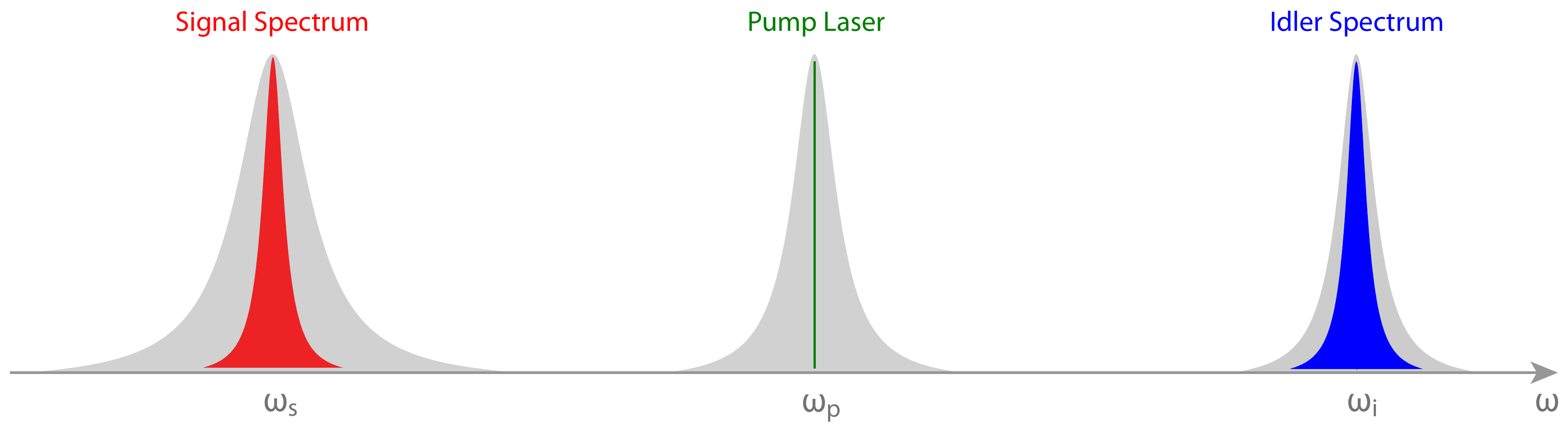}
\caption{\label{FigS4} {\bf Schematic of the identical spectra profile of signal and idler photons.} The gray Lorentzian functions represent cavity modes of involved optical modes. Pump laser has a narrow linewidth indicated by green line. Although signal mode has a broader loaded cavity linewidth than idler mode, the emitted signal (red) and idler (blue) photons share the same spectra profile due to energy conservation.}
\end{figure}

Remarkably, equation ~(\ref{Eq_SelfCor}) shows that the signal and idler photons exhibit identical self correlation, even when their individual cavity modes have different lifetimes. Figure~\ref{SelfCorrelation}(c) confirms this prediction in the time domain, with a direct comparison between $g_{\rm ss}^{(2)}(\tau)$ and $g_{\rm ii}^{(2)}(\tau)$. The identical self correlation agrees with the prediction of biphoton spectra in the frequency domain (Eq.~(\ref{Eq_Spectra})), where the energy conservation of the FWM process restricts signal and idler photons to exhibit identical emission spectra profile. As illustrated in Fig.~\ref{FigS4}, although the signal and idler cavity modes of the passive cavity have different cavity Qs and thus different linewidths, the emitted photons have the same spectra profile located around the signal (or idler) frequency because of the energy conservation.

Of particular interest is that Eq.~(\ref{Eq_SelfCor}) reveals that the self correlation function does not show a pure exponential function, in contrary to the common adoption of exponential fittings for the self correlation functions \cite{Morandotti14, Ramelow15, Benson11, Leuchs13, Thew14, Leuchs15}. Eq.~(\ref{Eq_SelfCor}) also suggests that self correlation function has a much broader linewidth than cross correlation function. The non-exponential and linewidth-broadening features can be seen more clearly when the signal and idler photons share the same lifetime ($\Gamma_{\rm ts} = \Gamma_{\rm ti} = \Gamma_t$). In this case, Eq.~(\ref{Eq_SelfCor}) reduces to $g_{\rm jj}^{(2)}(\tau) = (1 + \Gamma_t |\tau|/2)^2 e^{-\Gamma_t |\tau|}+1$. These features are observed experimentally by comparing Fig.~\ref{SelfCorrelation}(c) with Fig.~\ref{CrossCorrelation}(c). Clearly, $g_{\rm jj}^{(2)}(\tau)$ shows a waveform significantly broader than $g_{\rm si}^{(2)}(\tau)$. Although Eq.~(\ref{Eq_SelfCor}) is obtained here for a FWM process, it is universal to a cavity enhanced nonlinear process, regardless of FWM or SPDC \cite{Drummond90}.

\section{Heralded single photons}
\begin{figure}[t!]
\centering\includegraphics[width=1.0\columnwidth]{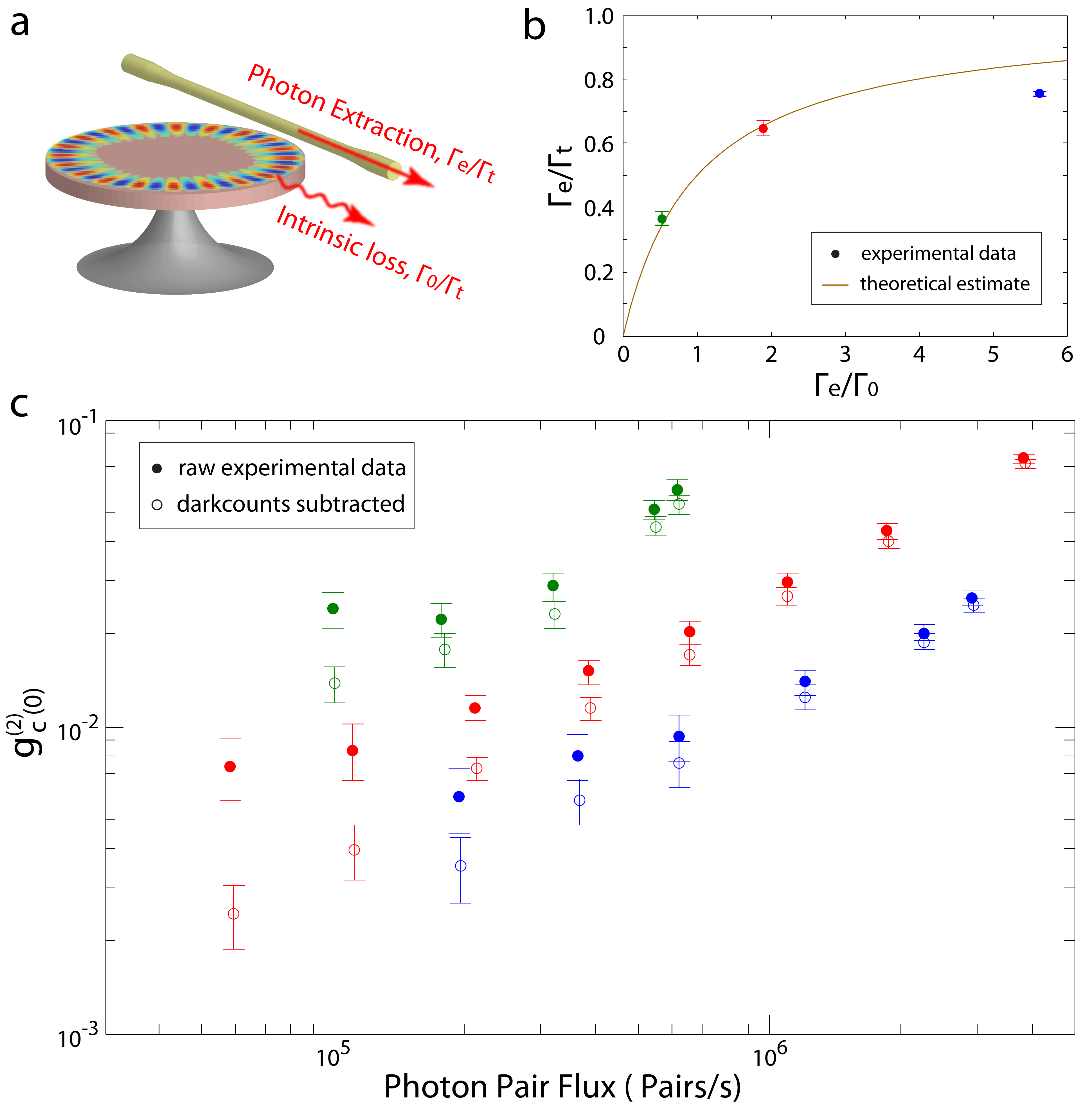}
\caption{\label{Fig7} {\bf Heralded single photons and conditional self correlations under various fiber-device coupling conditions.} (a) A schematic illustration of the photon extraction efficiency of the generated photons, where photons lost in the cavity can not be heralded. (b) By tuning the fiber-device distance, the external taper coupling rate is increased to improve the photon extraction efficiency. The yellow curve is the theoretical curve deduced from cavity transmission traces. (c) The heralded single photon source is characterized by the Hanbury-Brown-Twiss experiment. The data points have the coupling conditions described in (b), labeled by different colors. The solid circles represent the raw data without any corrections. The empty circles represent the processed data with detector darkcounts subtracted.}
\end{figure}

The high cross correlation and the single mode nature of the photons readily imply the application for heralding high-purity single photon Fock state. Heralding efficiency is an important figure of merit in heralded single photon sources \cite{Kartik12, Walmsley13, Eggleton13_2, Silberhorn13, Eggleton13, Morandotti14}. A critical parameter distinct in the cavity quantum electrodynamic system is photon extraction efficiency $\eta_{E}$. After a biphoton are created inside the cavity, individual photons can either be extracted out of the cavity into the coupling waveguide with an external coupling rate of $\Gamma_{e}$, or be lost inside the cavity at an intrinsic photon decay rate of $\Gamma_0$ due to the intrinsic material absorption or device scattering (Fig.~\ref{Fig7}(a)). Because only the extracted photons can be detected and heralded, the photon extraction efficiency of the heralded photons in the coupling waveguide is given by $\eta_E = \Gamma_e/\Gamma_t$.

In practice, as the signal mode at 1534.4~nm has an external coupling rate higher than the idler mode at 1497.1~nm (see Fig.~\ref{PL}(b)), we use the detection of an idler photon to herald the arrival of a signal photon. The flexible external coupling in our system results in a widely tunable photon extraction efficiency, as shown in Fig.~\ref{Fig7}(b), where a maximum $\eta_E$ of 76\% is achieved by over coupling the device. To characterize the performance of photon sources, preparation efficiency \cite{Walmsley13} is defined as the heralding efficiency in the single mode fiber, with the effects of optical components and detectors excluded. At the above coupling condition, the preparation efficiency is 51\%. The corresponding Klyshko efficiency \cite{Klyshko80} (i.e., raw heralding efficiency) is 3.4\%, which is primarily due to the low detection efficiency of the InGaAs SPDs (15\%) and the loss in optical components (-3.55~dB). The Klyshko efficiency can be improved by using detectors with higher efficiencies and optimizing the fiber-device coupling conditions.

The quality of heralded single photons can be characterized by the conditional self-correlation $g_c^{(2)}(0) \equiv \frac{\ave{E_s^\dag(t) E_s^\dag(t) E_s(t) E_s(t)}_i} {\ave{E_s^\dag(t)E_s(t)}_i^2 }$, where the subscript on $\ave{}_i$ denotes the condition of detecting an idler photon \cite{Grangier86}. The value of $g_c^{(2)}(0)$ describes the probability of heralding a multi-photon state. For an ideal heralded single photon Fock state, $g_c^{(2)}(0) =0$ indicates a perfect anti-bunching feature. To characterize the photon anti-bunching quality of the heralded single photons, we perform a Hanbury-Brown-Twiss experiment to measure the conditional self correlation (Fig.~\ref{Setup}(d)). The signal photons are separated into two channels through a 50:50 beam splitter, and the triple coincidence is detected among the idler channel and the two split signal channels. The conditional self correlation of the heralded single photon is thus given by \cite{Grangier86, Beck07},
\begin{eqnarray}
g_c^{(2)}(0) = \frac{N_{is_1s_2}N_{i}}{N_{is_1}N_{is_2}}, \label{conditionalselfcorrelaiton}
\end{eqnarray}
where $N_{is_1s_2}$ is the triple coincidence rate, $N_i$ is the counting rate of the idler photons, and $N_{is_1}/N_{is_2}$ are the double coincidence rates between the idler photon and the single photon in the two split channels, respectively.

Figure ~\ref{Fig7}(c) shows the recorded ${\rm g_{c}^{(2)}(0)}$ versus biphoton flux in three different coupling conditions, whose photon extraction efficiencies are shown in Fig.~\ref{Fig7}(b). ${\rm g_{c}^{(2)}(0)}$ remains below 0.08 over the entire recorded range of photon flux up to $\rm 3.9\times10^6~pairs/s$, with a lowest value of ${\rm 0.0059\pm0.0014}$ (${\rm 0.0035\pm0.0008}$) recorded at a pair flux of ${\rm 1.95 \times 10^5~pairs/s}$, without (with) subtracting detector dark counts. Moreover, for a certain photon pair flux, the conditional self correlation decreases as photon extraction efficiency increases. At the same photon-pair flux, a higher photon extraction efficiency results in a reduced photon numbers inside the cavity, which in turn reduces the probability of multi-photon generation. At a photon extraction efficiency of 76\% (blue dots in Fig.~\ref{Fig7}(c)), ${\rm g_{c}^{(2)}(0)}$ becomes ${\rm 0.026\pm0.001}$ at a pair flux of ${\rm 2.93 \times 10^6~pairs/s}$. To the best of our knowledge, these values are among the lowest ${\rm g_{c}^{(2)}(0)}$ at corresponding biphoton fluxes reported to date \cite{Walmsley13, Silberhorn13}, which demonstrate the superior quality of the heralded single photon Fock state produced in our device.

\section{Discussion}
In summary, we demonstrate the nonclassical nature of biphotons produced by cavity-enhanced FWM on a silicon chip, with a cross correlation of $\rm g_{si}^{(2)}(0) = (2.58 \pm 0.16) \times 10^4$, the highest value reported to date \cite{Kartik12, Walmsley13, Eggleton13_2, Silberhorn13, Eggleton13, Morandotti14, Sharping06, Takesue07, Fejer07, Baet09, Takesue10, Eggleton11, Bajoni12, Matsuda12, Eggleton12, Helmy12, Olislager13, Notomi13, Engin13, Mookherjea13, Kivshar14, Ducci14, Huang14, Peng14, Takesue14, Grassani15, Suo15, Zhu14, Thompson14, Wakabayashi15, Jiang15, Ramelow15, Savanier15, Ou99, Kobayashi04, Wong06, Polzik07, Mitchell08, Guo08, Scholz09, Benson11, Harris12, Leuchs13, Riedmatten13, Thew14, Leuchs15}. The generated photons are essentially in a single mode, which is confirmed with self-correlation of $\rm g_{ss}^{(2)}(0) = 1.90 \pm 0.05$ and $\rm g_{ii}^{(2)}(0) = 1.87 \pm 0.06$, respectively. The demonstrated photon cross-correlation violates the classical Schwartz inequality, ${\rm g_{si}^{(2)}(0) \le [g_{ss}^{(2)}(0) g_{ii}^{(2)}(0)}]^{1/2}$, by four orders of magnitude. The device has a spectral brightness of $\rm 1.06 \times 10^9 ~ pairs/s/mW^2/GHz$, which is the highest in current FWM-based devices \cite{Kartik12, Walmsley13, Eggleton13_2, Eggleton13, Morandotti14, Sharping06, Takesue07, Baet09, Takesue10, Eggleton11, Bajoni12, Matsuda12, Eggleton12, Engin13, Olislager13, Notomi13, Mookherjea13, Huang14, Peng14, Takesue14, Grassani15, Suo15, Wakabayashi15, Ramelow15, Jiang15, Thompson14, Kumar04}. In particular, we observe the waveform asymmetry of the biphoton cross correlation and revealed the identity and non-exponential nature of the self correlation functions of individual signal and idler photons. The strong photon correlation and high efficiency of biphoton production enable us to herald single photons with conditional self correlation $\rm g_{c}^{(2)}(0)$ as low as $\rm 0.0059 \pm 0.0014$ at a pair flux of $\rm 1.95 \times 10^5$~pairs/s, which remains below $\rm 0.026 \pm 0.001$ for biphoton flux up to $\rm 2.93 \times 10^6$~pairs/s, with a photon preparation efficiency in the single-mode fiber up to 51\%. The recorded ${\rm g_c^{(2)}(0)}$ values are among the best values reported \cite{Walmsley13, Silberhorn13}.

The high quality, high efficiency, and single-mode nature of biphotons and heralded single photons produced directly in the telecom band, together with the CMOS compatibility of the device platform, render our device a promising source for a variety of integrated quantum photonic applications including quantum key distribution, hybrid macroscopic-microscopic entanglement, and intra-/inter-chip quantum state teleportation. The exploration of the detailed waveform structure of cross- and self- photon correlations advance our understanding of biphoton statistics of cavity-enhanced nonlinear optical processes, which may have profound impact on quantum functionalities such as quantum frequency conversion and light-matter interfacing that require detailed information of photon wavepackets. There has been significant development in the past few years on various chip-scale platforms for photon generation \cite{Kartik12, Walmsley13, Eggleton13_2, Silberhorn13, Eggleton13, Morandotti14,Sharping06, Takesue07, Fejer07, Baet09, Takesue10, Eggleton11, Bajoni12, Matsuda12, Eggleton12, Helmy12, Olislager13, Notomi13, Engin13, Mookherjea13, Kivshar14, Ducci14, Huang14, Peng14, Takesue14, Grassani15, Suo15, Zhu14, Thompson14, Wakabayashi15, Jiang15, Ramelow15, Savanier15}. Our work unambiguously demonstrates that silicon photonic chips are superior material and device platforms for integrated quantum photonics. Further development of quantum photonic functionalities based upon our devices would form a fundamental building block towards an integrated quantum photonic interconnect.

\section*{Acknowledgments}
We thank Stefan Preble for the loan of superconducting nanowire single-photon detectors. We thank Oskar Painter for helpful discussions. This work is supported by National Science Foundation (NSF) under grant No.~ECCS-1351697. The device fabrication was performed in part at the Cornell NanoScale Science \& Technology Facility (CNF), a member of the National Nanotechnology Infrastructure Network.\\

\section*{Appendix}
\begin{figure}[t!]
\includegraphics[width=1.0\columnwidth]{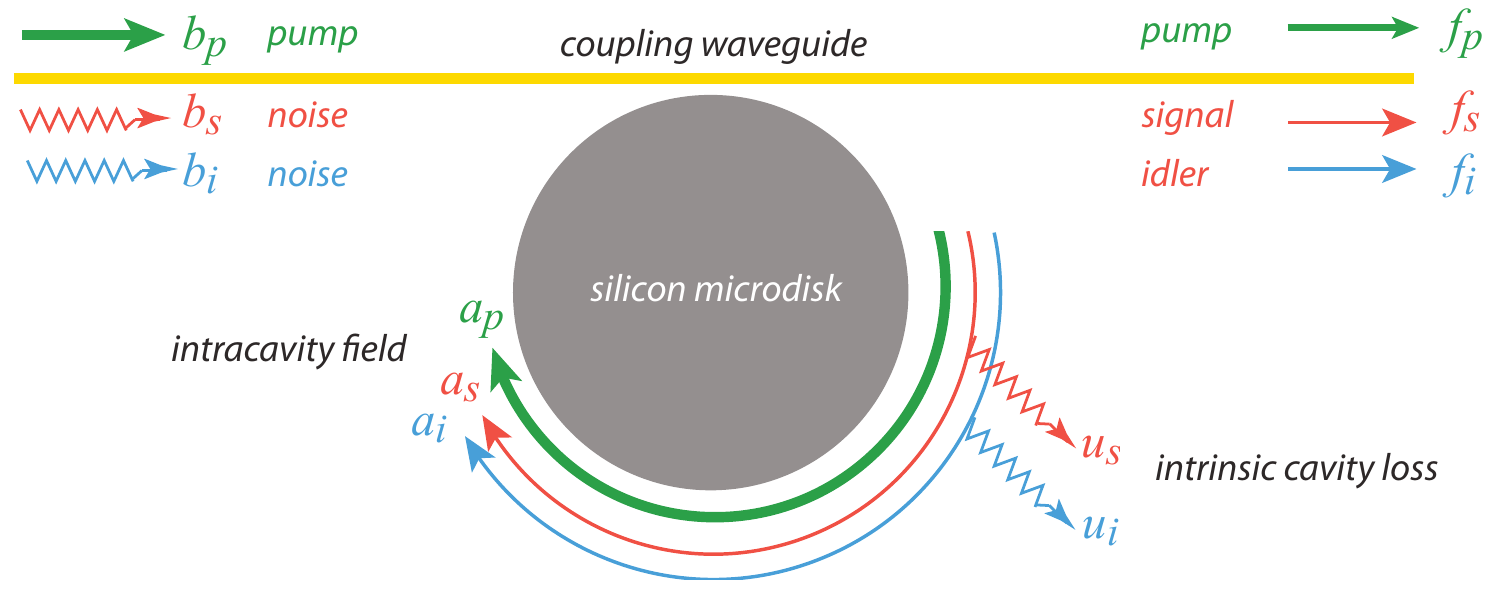}
\caption{\label{FigS1} Schematic of biphoton generation in a silicon microdisk resonator. }
\end{figure}

In this section, we provide a theoretical description of biphoton generation through spontaneous degenerate FWM in the triply-resonant cavity. We only consider Kerr nonlinear optical interaction, since two-photon absorption and free-carrier effect are negligible for the optical power employed in our system. As shown in Fig.~\ref{FigS1}, a pump wave is launched into the resonator to generate signal and idler photon pairs, which are coupled out of the cavity into the delivery waveguide. The dynamics among the three cavity modes at frequencies $\omega_{\rm 0j}$ ($j=p,s,i$) can be described by the following equations of motion \cite{Jiang15},
    \begin{widetext}
\begin{eqnarray}
\frac{da_p}{dt} &=& (-i\omega_{\rm 0p} - \Gamma_{\rm tp}/2) a_p + i g a_p^\dag a_p^2 + i \sqrt{\Gamma_{\rm ep}} b_p(t)+ i \sqrt{\Gamma_{\rm 0p}} u_p(t), \label{dapdt} \\
\frac{da_s}{dt} &=& (-i\omega_{\rm 0s} - \Gamma_{\rm ts}/2) a_s + 2 i g a_p^\dag a_p a_s + i g a_i^\dag a_p^2 + i \sqrt{\Gamma_{\rm es}} b_s(t) + i \sqrt{\Gamma_{\rm 0s}} u_s(t), \label{dasdt} \\
\frac{da_i}{dt} &=& (-i\omega_{\rm 0i} - \Gamma_{\rm ti}/2) a_i + 2 i g a_p^\dag a_p a_i + i g a_s^\dag a_p^2 + i \sqrt{\Gamma_{\rm ei}} b_i(t) + i \sqrt{\Gamma_{\rm 0i}} u_i(t), \label{daidt}
\end{eqnarray}
    \end{widetext}
where the intracavity field operator $a_j$ ($j=p,s,i$) is normalized such that $a_j^\dag a_j$ represents the photon number operator. $\Gamma_{\rm 0j}$ and $\Gamma_{\rm tj} = \Gamma_{\rm 0j} + \Gamma_{\rm ej}$ ($j=p,s,i$) are the photon decay rates of the intrinsic and loaded cavity, respectively. $b_j$ is the field operator of the incoming wave at carrier frequency $\omega_j$ inside the coupling waveguide, which is normalized such that $b_j^\dag b_j$ represents the input photon flux. $b_j$ satisfies the commutation relation of $[b_j(t), b_j^\dag (t')] = \delta (t-t')$. $u_j$ is the noise operator associated with the intrinsic cavity loss, which satisfies the commutation relation of $[u_j(t), u_j^\dag (t')] = \delta (t-t')$. ${\rm g=c \eta n_2 \hbar \omega_p \sqrt{\omega_s \omega_i}/(n_s n_i \bar{V})}$ describes the vacuum coupling rate of the FWM process, where ${\rm \eta}$, ${\rm n_2}$, and ${\rm \bar{V}}$ are the spatial overlap among the interacting modes, the Kerr nonlinear coefficient, and the effective mode volume, respectively. In the theoretical prediction, the mode overlap (${\rm \eta}$) is unity since the three optical modes share the same spatial mode profile. The third-order nonlinearity is ${\rm n_2 = 2.23 \times 10^{-5}~cm^2/GW}$. The effective volume of the employed optical modes is ${\rm \bar{V} = 9.01 \mu m^3}$, simulated by finite-element method.

The pump mode is launched with a continuous-wave laser and can be treated as a classical field. For spontaneous FWM process, the signal and idler modes stay at single photon regime, and the pump mode can be assumed non-depleted. Making a Fourier transform as $F_j(\omega)=\int_{-\infty}^{+\infty} e^{i(\omega-\omega_j) t} F_j(t) dt$, where $F_j$ stands for operator $a_j$, $b_j$, or $u_j$ ($j=p,s,i$), we can solve Eqs.~(\ref{dapdt})-(\ref{daidt}) analytically in the frequency domain to obtain
\begin{eqnarray}
a_p(\omega) &=& i \sqrt{\Gamma_{\rm ep}} b_p(\omega)/(\Gamma_{\rm tp}/2 - i \omega), \label{ap} \\
a_s(\omega) &=& A(\omega) n_s(\omega) +B(\omega) n_i^\dag(-\omega), \label{as}\\
a_i(\omega) &=& A(\omega) n_i(\omega) +B(\omega) n_s^\dag(-\omega), \label{ai}
\end{eqnarray}
where the noise operators $n_s$ and $n_i$ are given by
\begin{eqnarray}
n_s(\omega) &\equiv& i \sqrt{\Gamma_{\rm es}} b_s(\omega) + i \sqrt{\Gamma_{\rm 0s}} u_s(\omega), \label{ns} \\
n_i(\omega) &\equiv& i \sqrt{\Gamma_{\rm ei}} b_i(\omega) + i \sqrt{\Gamma_{\rm 0i}} u_i(\omega). \label{ni}
\end{eqnarray}
In Eqs.~(\ref{as}) and (\ref{ai}), $A(\omega)$ and $B(\omega)$ have the following expressions,
\begin{eqnarray}
A(\omega) &=& \frac{\Gamma_{\rm ti}/2 - i\omega}{(\Gamma_{\rm ts}/2 - i\omega)(\Gamma_{\rm ti}/2 - i\omega) - (g N_p)^2}, \label{eq_A}\\
B(\omega) &=& \frac{-i g a_p^2 }{(\Gamma_{\rm ts}/2 - i\omega)(\Gamma_{\rm ti}/2 - i\omega) - (g N_p)^2}, \label{eq_B}
\end{eqnarray}
where $N_p = \ave{a_p^\dag a_p}$ is the average photon number of the pump wave inside the cavity. Equations (\ref{eq_A}) and (\ref{eq_B}) include the multi-photon generation induced by the stimulated FWM. In the single-photon regime, where $g N_p \ll (\Gamma_{\rm ts}~ {\rm and}~ \Gamma_{\rm ti})$, these equations reduce to
\begin{eqnarray}
A(\omega) &\approx& \frac{1}{\Gamma_{\rm ts}/2 - i\omega}, \label{eq_Ar}\\
B(\omega) &\approx& \frac{-i g a_p^2 }{(\Gamma_{\rm ts}/2 - i\omega)(\Gamma_{\rm ti}/2 - i\omega)}. \label{eq_Br}
\end{eqnarray}

As the transmitted field for each photon mode is given by \cite{MilburnBook}
\begin{eqnarray}
f_j = b_j + i \sqrt{\Gamma_{\rm ej}} a_j, \label{f_j}
\end{eqnarray}
Eqs.~(\ref{ap})-(\ref{f_j}) can be applied to find the photon statistics. For example, the spectra of the signal and idler photons are given by
\begin{eqnarray}
S_s(\omega) &\equiv& \ave{f_s^\dag (\omega) f_s(\omega)} = \Gamma_{\rm es} \Gamma_{\rm ti} \left| B(\omega) \right|^2,  \label{S_s} \\
S_i(\omega) &\equiv& \ave{f_i^\dag (\omega) f_i(\omega)} = \Gamma_{\rm ei} \Gamma_{\rm ts} \left| B(\omega) \right|^2, \label{S_i}
\end{eqnarray}
which shows that the signal and idler exhibits an identical spectral profile, a result of energy conservation for the FWM process,
\begin{eqnarray}
S_j(\omega) \propto \left| B(\omega) \right|^2 = \frac{g^2 N_p^2}{[\omega^2+(\Gamma_{\rm ts}/2)^2][\omega^2+(\Gamma_{\rm ti}/2)^2]}, \label{S_j}
\end{eqnarray}
which is Eq.~(\ref{Eq_Spectra}) in the main text. Interestingly, Eq.~(\ref{S_j}) also shows that the emission spectrum of signal/idler photons exhibits a spectral width narrower than that of the passive cavity mode. This spectral narrowing arises from the energy conservation among interacting cavity modes, regardless of SFWM or SPDC processes \cite{Drummond90, Sipe15}. At the same time, the emitted photon fluxes of individual signal and idler are given by
\begin{eqnarray}
\ave{f_s^\dag (t) f_s(t)} &=& \frac{\Gamma_{\rm es} \Gamma_{\rm ti}}{2\pi} \int_{-\infty}^{+\infty}{\left| B(\omega) \right|^2 d\omega} \nonumber \\
&\approx& \frac{2 \Gamma_{\rm es} (g N_p)^2}{\Gamma_{\rm ts} \bar{\Gamma}}, \label{SignalFlux}\\
\ave{f_i^\dag (t) f_i(t)} &=& \frac{\Gamma_{\rm ei} \Gamma_{\rm ts}}{2\pi} \int_{-\infty}^{+\infty}{\left| B(\omega) \right|^2 d\omega} \nonumber \\
&\approx& \frac{2 \Gamma_{\rm ei} (g N_p)^2}{\Gamma_{\rm ti} \bar{\Gamma}}. \label{IdlerFlux}
\end{eqnarray}

In particular, the cross correlation between the signal and idler photons is found to be
\begin{eqnarray}
g_{\rm si}^{(2)}(\tau) & \equiv & \frac{\ave{f_{\rm i}^\dag (t) f_{\rm s}^\dag (t+\tau) f_{\rm s} (t+\tau) f_{\rm i}(t)}}{\ave{f_{\rm s}^\dag (t) f_{\rm s}(t)} \ave{f_{\rm i}^\dag (t) f_{\rm i}(t)}}, \nonumber \\
& = & \frac{\left| \int_{-\infty}^{+\infty} B(-\omega)[\Gamma_{\rm ts} A(\omega)-1]e^{-i \omega \tau} d \omega \right|^2}{ \Gamma_{\rm ts} \Gamma_{\rm ti} \left| \int_{-\infty}^{+\infty} \left| B(\omega) \right|^2 d \omega  \right|^2}+1, \nonumber \\
& = &
  \begin{cases}
    \frac{\Gamma_{\rm ts} \Gamma_{\rm ti}}{4 (g N_{\rm p})^2} e^{-\Gamma_{\rm ts} \tau} + 1 & \quad (\tau \ge 0) \\
    \frac{\Gamma_{\rm ts} \Gamma_{\rm ti}}{4 (g N_{\rm p})^2} e^{\Gamma_{\rm ti} \tau}  + 1 & \quad (\tau < 0)
  \end{cases}, \label{gsi}
\end{eqnarray}
which is Eq.~(\ref{Eq_CrossCor}) in the main text. Therefore, the emission flux of correlated biphotons is given by
\begin{eqnarray}
R_c &=& \ave{f_{\rm s}^\dag (t) f_{\rm s}(t)} \ave{f_{\rm i}^\dag (t) f_{\rm i}(t)} \int_{-\infty}^{+\infty} {[g_{\rm si}^{(2)}(\tau) -1] d\tau} ,\nonumber \\
&=& \frac{\Gamma_{\rm es} \Gamma_{\rm ei}}{\Gamma_{\rm ts} \Gamma_{\rm ti}} \frac{2 (g N_p)^2}{\bar{\Gamma}}. \label{R_c}
\end{eqnarray}

\begin{figure*}[t!]
\centering\includegraphics[width=1.5\columnwidth]{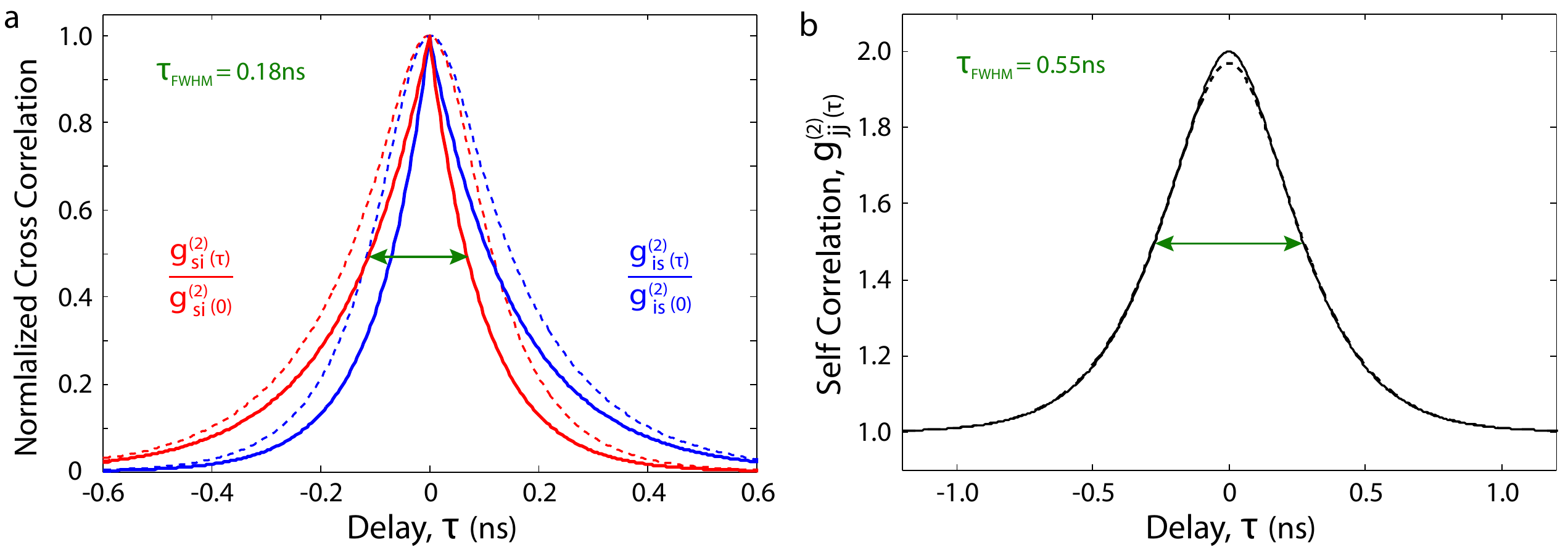}
\caption{\label{FigS2} {\bf Theoretical cross/self correlations.} (a) Normalized cross correlations without the broadening from instrument response function (IRF) show asymmetric exponential decay in solid red and blue lines. The green arrow specifies a full-wave-half-maximum (FWHM) time of 0.18 ns. (b) Self correlations without IRF show the identical and non-exponential spectral profile for both signal and idler photons in the solid black line. The green arrow specifies a FWHM time of 0.55 ns, which is much broader than the FWHM time in cross correlations. The dashed lines in (a) and (b) correspond to the predictions of cross and self correlations with IRF in Figs.~\ref{CrossCorrelation}(c) and \ref{SelfCorrelation}(c), respectively.}
\end{figure*}
On the other hand, the self correlations of signal or idler photons are found to be
\begin{eqnarray}
g_{\rm jj}^{(2)}(\tau) & \equiv & \frac{\ave{f_{\rm j}^\dag (t) f_{\rm j}^\dag (t+\tau) f_{\rm j} (t+\tau) f_{\rm j}(t)}}{\ave{f_{\rm j}^\dag (t) f_{\rm j}(t)}^2}, \nonumber \\
& = & \frac{\left| \int_{-\infty}^{+\infty} {\left| B(\omega) \right|}^2  e^{-i \omega \tau} d \omega  \right|^2}{\left| \int_{-\infty}^{+\infty} {\left| B(\omega) \right|}^2 d \omega  \right|^2}+1, \nonumber \\
& = & \left[ \frac{\Gamma_{\rm ts} e^{-\Gamma_{\rm ti}|\tau|/2} - \Gamma_{\rm ti} e^{-\Gamma_{\rm ts}|\tau|/2} }{\Gamma_{\rm ts} - \Gamma_{\rm ti}}\right]^2 +1, \label{gjj}
\end{eqnarray}
which is Eq.~(\ref{Eq_SelfCor}) in the main text. The signal and idler photons exhibit identical self correlation profile, regardless of the potential Q difference of their passive cavity modes. When the signal and idler mode exhibits a same photon decay rate, $\Gamma_{\rm ts} = \Gamma_{\rm ti} \equiv \Gamma_t$, Eq.~(\ref{gjj}) reduces to a simple form of $g_{\rm jj}^{(2)}(\tau) = (1 + \Gamma_t |\tau|/2)^2 e^{-\Gamma_t |\tau|}+1$, which shows clearly the non-exponential nature of the self correlation.

In practice, the coincidence counting system generally exhibits a finite instrument response primarily due to the timing jitters of the single photon detectors (gray dots in Fig.~\ref{CrossCorrelation}). As a result, the experimentally recorded photon correlations are the convolution between Eq.~(\ref{gsi}) (or Eq.~(\ref{gjj})) and the instrument response function (IRF) of the coincidence counting system. Fig.~\ref{FigS2} shows this effect, where the dashed lines include the instrument response, compared with the ideal case of Eqs.~(\ref{gsi}) and (\ref{gjj}) (solide lines). The full width at half maximum (FWHM) of the IRF of our detection system is about 106 ps. It broadens the cross correlation function by a certain extent (Fig.~\ref{FigS2}(a)) but its impact on the self correlation is small, only decreasing its peak value by a small amount from 2 to 1.98. In particular, Fig.~\ref{FigS2}(a) verifies the waveform asymmetry of the cross correlation. Fig.~\ref{FigS2}(b) shows the non-exponential profile for self correlation, whose FWHM time (0.55 ns) is much broader than the FWHM time of cross correlation (0.18 ns). As discussed above, the self correlations are identical for signal and idler photons (Fig.~\ref{FigS2}(b)) and they share the same emission spectra profile(Eqs.~(\ref{S_s})-(\ref{S_j})).

\end{document}